\documentclass[preprint,showkeys,longbibliography]{revtex4-1}
\usepackage{graphicx}                         
\usepackage{dcolumn}                          
\usepackage[version=3]{mhchem}                
\usepackage[T1]{fontenc}                      

\usepackage{epsfig}
\usepackage{amsmath}

\usepackage{epstopdf}
\usepackage{amsbsy}
\usepackage{multirow}
\usepackage{bm}
\usepackage{caption}
\usepackage{siunitx}
\usepackage{CJK}
\usepackage{textcomp}
\usepackage{comment}
\usepackage{soul}
\usepackage[ruled,vlined]{algorithm2e}
\usepackage{caption}
\usepackage{subcaption}
\usepackage{array}
\usepackage{comment}
\usepackage{amsfonts}
\usepackage[table]{xcolor}

\usepackage{multirow}

\newcolumntype{P}[1]{>{\centering\arraybackslash}p{#1}}

\begin{document}

\title{Actuation mechanisms in twisted and coiled polymer actuators using finite element model}
\author{Gurmeet Singh$^{1,2}$}
\email{Corresponding author. 
\newline \mbox{\textit{E-mail address}: gmsingh@umich.edu (G. Singh).}}
\author{Qiong Wang$^{3}$}
\author{Samuel Tsai$^{3}$}
\author{Sameh Tawfick$^{3}$}
\author{Umesh Gandhi$^{2}$}
\author{Veera Sundararaghavan$^{1}$}

\affiliation{$^1$Department of Aerospace Engineering, University of Michigan, Ann Arbor, MI, USA 48109}
\affiliation{$^2$Future Mobility Research Department, Toyota Research Institute North America, Ann Arbor, MI, USA, 48105}
\affiliation{$^3$Department of Mechanical Science and Engineering, University of Illinois Urbana-Champaign, Urbana, IL, USA, 61801}

\begin{abstract}
Twisted and coiled polymer actuators (TCPAs) offer the advantages of large stroke and large specific work as compared to other actuators. There have been extensive experimental investigations towards understanding their actuation response, however, a computational model with full material description is not utilized to probe into the underlying mechanisms responsible for their large actuation. In this work, we develop a three-dimensional finite element model that includes the physics of the fabrication process to simulate the actuation of TCPA under various loading and boundary conditions. The model is validated against the experimental data and used to explore the factors responsible for actuation under free and isobaric conditions. The model captures the physics of the angle of twist in the fiber and the distinction between the homochiral and heterochiral nature of TCPA actuation response. The simulations show that the anisotropy in the thermal expansion coefficient (CTE) matrix plays a major role in large actuation irrespective of the anisotropy or isotropy in the elasticity tensor. We further investigate the extent of anisotropy in thermal expansion and the parametric studies show that the key for TCPA actuation is the absolute value of mismatch in thermal expansion even if the material has positive or negative CTE in both directions of the fiber. Furthermore, we propose a new shell-core composite-based TCPA concept by combining the epoxy and hollow Nylon tubes to suppress the creep in TCPA. The results show that the volume fraction of epoxy-core can be tuned to attain a desired actuation while offering a stiffer and creep-resistant response. This framework provides a wider application for probing various kinds of TCPAs and enhancing their actuation performance.
\end{abstract}

\keywords{Finite element modeling, Actuation, Semi-crystalline polymers, TCPA, Material anisotropy}

\maketitle

\section{Introduction}

Twisted and coiled polymer actuators (TCPAs) are a type of artificial muscle that works by external stimuli of various forms: electrical, thermal, and change in pH, etc.\cite{haines2014artificial,haines2016new,tawfick2019stronger,kongahage2019actuator}. These are fabricated from polymer fibers by twisting and coiling them into a helix, as a result, they can actuate axially or twist upon the application of stimulus. These actuators have demonstrated impressive performance in terms of large actuation stroke ($>50\%$), power density (27.12 kW/kg), and specific work (2.48 kJ/kg)\cite{wu2020position,zhou2021power}. These capabilities have attracted attention to the potential applications of TCPAs in smart materials, soft robotics, and artificial muscles\cite{Saharan2019,tawfick2019stronger}. Additionally, TCPAs offer several advantages over conventional actuation methods, such as high contraction ability, high work output, low-cost materials, and simple fabrication, which make them suitable for applications in biomedical devices, and soft robotics\cite{haines2016new,tawfick2019stronger,chu2021unipolar,almubarak2017twisted}.

TCPAs offer large and gradual actuation strokes as opposed to the shape-memory alloys(SMAs), which show smaller and sudden strokes due to rapid phase transformation\cite{liang1992design,kheirikhah2011review,elahinia2016shape,kim2023shape}. These actuators are primarily fabricated from semicrystalline polymer mono-filament fibers, twisted and coiled in a particular way to create a helical coiled structure. These actuators can contract or extend during actuation, depending on the twist direction in reference to the coil axis, and cooling or heating w.r.t a reference temperature. However, due to their soft nature and therefore, being prone to axial buckling, these actuators are more useful under contraction than in extension. Despite a promising scope of potential applications of TCPAs in various fields, there are also some key challenges associated with these actuators that limit their utilization in critical applications, such as their relatively slow response time and difficulty in controlling their complex motion, creep deformation, and sensitivity to moisture\cite{mirvakili2018artificial,zou2021progresses,leng2021recent}. Ongoing research is focused on developing new materials, heating and cooling strategies, and understanding their performance, design strategies, and alternative materials toward large actuation forces to improve the performance of TCPA muscles\cite{zou2021progresses,leng2021recent,leng2023tethering,lang2024emerging,liang2020comparative,higueras2021artificial}. Therefore, understanding the deformation mechanisms and role of material and geometrical architecture becomes of paramount interest in enhancing the performance and utilization of these actuators. Since the introduction of TCPAs, there has been significant interest in improving their performance by the choice of material, actuation methods, coil geometric parameters, and architecture of multiple fibers \cite{lang2024emerging}. Most of these investigations have been experimental which limits the scope of mechanistic understanding and makes it tedious to explore the design envelope of these actuators, considering the complex physics and coupling between various parameters\cite{tsabedze2021design,wu2017compact,chen2024effect}.

Developing a high-fidelity computational model with predictive capabilities is important towards understanding the actuation mechanisms, and, therefore, improve the performance metric of the TCPA actuators. A numerical model can help explore the design space of material parameters, fabrication processes, geometric parameters of the coil, and the nature of loading conditions (both thermal and mechanical). A computational model can be used not only for exploring the design space but also to gain insights into the underlying mechanisms that dictate the actuation of these actuators. There have been several coil kinematics models, energy-based models, and beam theory-based models developed\cite{pawlowski2018modeling,Wu2020}.  Pawloski et al. proposed a beam formulation-based model that includes the change in coil length as a function of temperature. However, such a model assumes the actuation vs. temperature response is known explicitly from experiments conducted under various conditions of the actuator as a whole. Zhang et al. developed a model for a supercoiled TCPA that captures the hysteresis which is a key aspect to capture for a TCPAs\cite{zhang2017modeling}. An analytical model by Karami et al. captures the multi-physics of the electroactuation via Joule heating and models the mechanical response of the TCPA using Kirchhoff–Love's beam theory\cite{karami2017modeling}. The majority of such beam formulations-based models lack the ability to capture the fiber-level finite strains and deflections that a TCPA undergoes. Towards including this aspect, a study by Sun et al., modeled the TCPA using Elastica Cosserat rod theory that captures the large deflections and rotations\cite{sun2021physics} using Euler's Elastic formulation. This model was also able to capture response of a conical TCPA. However, this model does not include any information on the microstructure at the cross-section level and fails to capture how the radially varying twist changes it and its effects on the actuation response. It primarily models the bulk-level behavior of the twisted fiber\cite{sun2021physics}. Wang et al. proposed a new model that captures the strain energy stored in crystalline and amorphous phases of the coil that provides the mechanistic understanding of the TCPA response due to interaction between the constituent phases\cite{WANG2024109440}. An energy-based semi-analytical formulation was proposed by Yang and Li that takes into consideration of the nano-scale deformation mechanisms of a continuum-scale model\cite{Yang2016TCPA}. It captures the twist by modeling the variable material orientation as a function of the radius by discretizing the cross-section into a finite number of concentric annuli. This captures the reorientation of the fiber microstructure due to the twist by computing the effective anisotropic elasticity tensor and the thermal expansion coefficient (CTE) matrix. It is a computationally efficient model that encapsulates the material non-linearity and spatial variation adequately. This model successfully includes the physics of microstructure, however, owing to its low dimensionality as a beam-based formulation, it lacks the full description of the geometry and the effect of deformations in the cross-section in the concentric annuli. This model takes in the effective stiffness of the cross-section and is agnostic to local cross-sectional deformations. Furthermore, its applications are limited to a simple coil-shaped geometry which is present in most cases, however, it fails to consider the non-linear finite deformations. Additionally, most of these models are not adequate for predicting the load-free actuation response which is a major limitation to investigate the mechanisms under free actuation response in TCPAs. Hunt et al. used Yang and Li's description to extend this work to a three-dimensional finite element (FE) model where they assigned the effective thermomechanical material properties to the entire cross-section\cite{Hunt2021}. Their model captures the actuation under different conditions well but does not consider the annular description of microstructure despite being a detailed 3D finite element (FE) model, instead assigns the effective cross-sectional stiffness and CTE properties. Additionally, the study did not investigated into the actuation mechanisms of the TCPAs.

In summary, the existing models use either curve-fitting of the actuation response to evaluate the coil model parameters or simply lack the detailed material and geometrical description that limits the extent of physics represented. A full three-dimensional finite element model with anisotropic material properties has not been developed and has not been used to probe into the actuation mechanisms of TCPAs, and which is the focus of this work. We use the multi-scale description of the twisted fiber proposed by Yang and Li\cite{Yang2016TCPA} and is discussed in the method section in detail. The present model utilizes the post-fabrication microstructure being a concentric helical laminate of variable orientation. The model is able to capture the load-free actuation response observed in TCPAs which is major limitation of beam-based one dimensions models. We use the transversely-isotropic thermal and elastic material properties of the precursor fiber. We use some material properties from literature, some from in-house testing to construct the transversely isotropic material stiffness tensor and CTE matrix. We verify the physics captured by the model by studying various cases such as the influence of chirality (homochiral vs. heterochiral), the influence of angle of twist angle, etc. We validated this model against the experimental test of a fabricated TCPA. Subsequently, we probe into the underlying actuation mechanisms of TCPAs, by investigating the effect of anisotropy in elastic stiffness tensor and CTE matrix of fiber material. Lastly, we present the simulation results of a core-shell composite concept to develop creep-resistant TCPAs. 

\section{Methods}

In this work, a microstructure-based finite element model is developed for the twisted and coiled polymer actuators. The drawing process of the semi-crystalline polymer leads to anisotropy in the precursor fibers used to fabricate TCPAs. These monofilament fibers are transversely isotropic in nature\cite{Chen1981,Choy1979,Choy1984,Chen1980,Choy1981,White1984,WANG2024109440,singh2023computational}. As a result of the drawing process, the crystalline phase (lamellae) aligns preferentially along the fiber axis\cite{Choy1981}. The distribution of orientation of crystalline phase is random that results in isotropy in the cross-section of the fiber\cite{Chen1981,Choy1979,Choy1981}. We solve the static equilibrium equation for a three-dimensional coil domain in the absence of any body force using finite element method in ABAQUS\cite{smith2009abaqus}:
\begin{align}
    \nabla.\mathbf{\sigma}&=0
\end{align}
    where, $\mathbf{\sigma}$ is the stress tensor at a material point. The constitutive law determines the stress tensor in terms of strain tensor at a material point and we assume a temperature dependent elastic response with transversely isotropic material stiffness tensor (detailed in Appendix).


\subsection{Concentric laminate model and material properties}\label{mat-prop-section}

\begin{figure}[h]
    \centering
    \includegraphics[trim={0.0cm 0cm 13cm 0.75cm},clip=true ,width=1.0\textwidth]{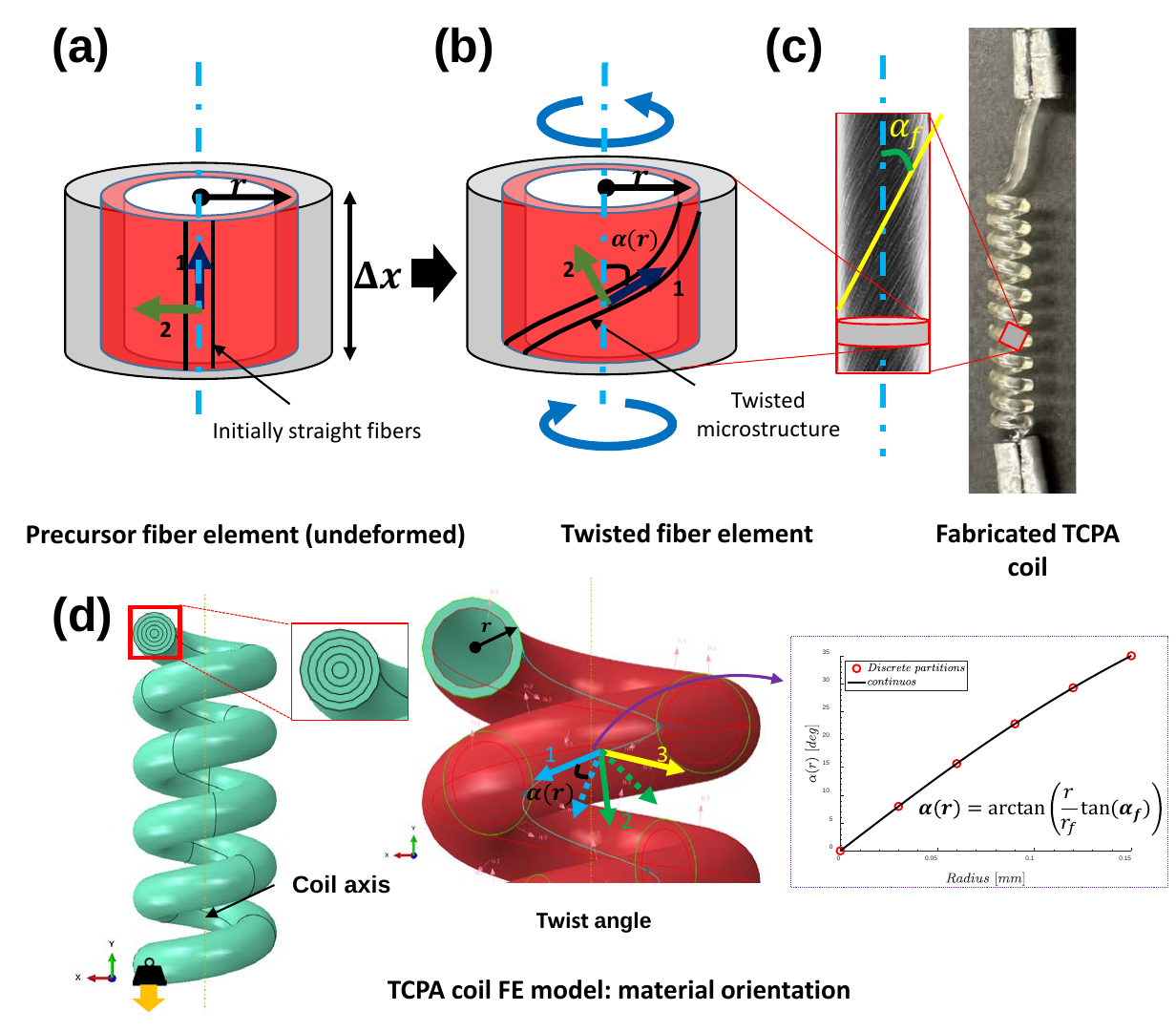}
    \caption{(a) Initially straight precursor fiber segment, (b) twisted state of semi-crystalline material fiber, (c) inserted twist ($\alpha_r$) in the fiber, (d) local material orientation assigned for concentric laminates in the TCPA FE model}
    \label{fig:twistSchematic}
\end{figure}

The TCPA muscles are fabricated by twisting a straight semi-crystalline fiber followed by coiling the twisted fiber by itself or around a mandrel \cite{haines2014artificial}. The drawn fibers have a micro-structure with the crystalline phase being highly oriented along the draw direction and randomly oriented in the fiber cross-section\cite{Choy1984}, i.e. isotropic in the radial direction of the fiber. The basis of the model in this work lies in the assumption that the microstructure rotates with the geometric twist introduced in the transversely isotropic precursor fiber during the fabrication process of the actuator\cite{Yang2016TCPA,haines2014artificial}. Consider an infinitesimal segment of the fiber in the precursor fiber (shown in Figure \ref{fig:twistSchematic}(a)) where the $axis-1$ and $axis-2$ represent the direction of transversely-isotropic mono-filament. Upon introducing the twist in this fiber, the crystalline phase subtends an angle with the fiber axis as shown in Figure \ref{fig:twistSchematic}(b). The reorientation angle varies with the radial position ($r$) of a material point  of the fiber as a geometric function\cite{Yang2016TCPA} :

\begin{equation}\label{alpha_eq}
    \alpha(r) = \text{arctan}\left(\frac{r}{r_f}\text{tan}(\alpha_f) \right)
\end{equation}

where $r_f$ is the fiber radius and $\alpha_f$ is the twist angle at fiber surface and this can be measured experimentally after the TCPA fabrication (refer inset of Figure \ref{fig:twistSchematic}(c)). Note that the angle of twist ($\alpha(r)$) is a continuous function of the radial location from the fiber axis. In the present FE model, we divide the fiber cross-section into finite number of concentric laminates with radially varying material orientation assigned using Eqn. \ref{alpha_eq}. Figure \ref{fig:twistSchematic}(d) shows one of such helical laminates with local material orientation rotated an angle ($\alpha(r)$).

The following key assumptions are made about the drawn fiber after it is twisted and coiled in a TCPA:
\begin{itemize}
    \item[-] No residual stress is present after the annealing of the TCPA.
    \item[-] The microstructure is not altered by the annealing during the fabrication process.
    \item[-] The amount of twist is a function of the radial location on the fiber and follows a geometric relation given by Eqn. \ref{alpha_eq}\cite{Yang2016TCPA,haines2014artificial}.
    \item[-] No creep occurs i.e. the constitutive model is assumed with no viscoelasticity response. Although Nylon-based TCPA shows viscoelastic behavior\cite{guo2025driving} but the focus of present study does not include time-dependent response.
\end{itemize}

\begin{figure}[h]
    \centering
    \includegraphics[trim={0.cm 3.5cm 18.0cm 0.0cm},clip=true ,width=0.9\textwidth]{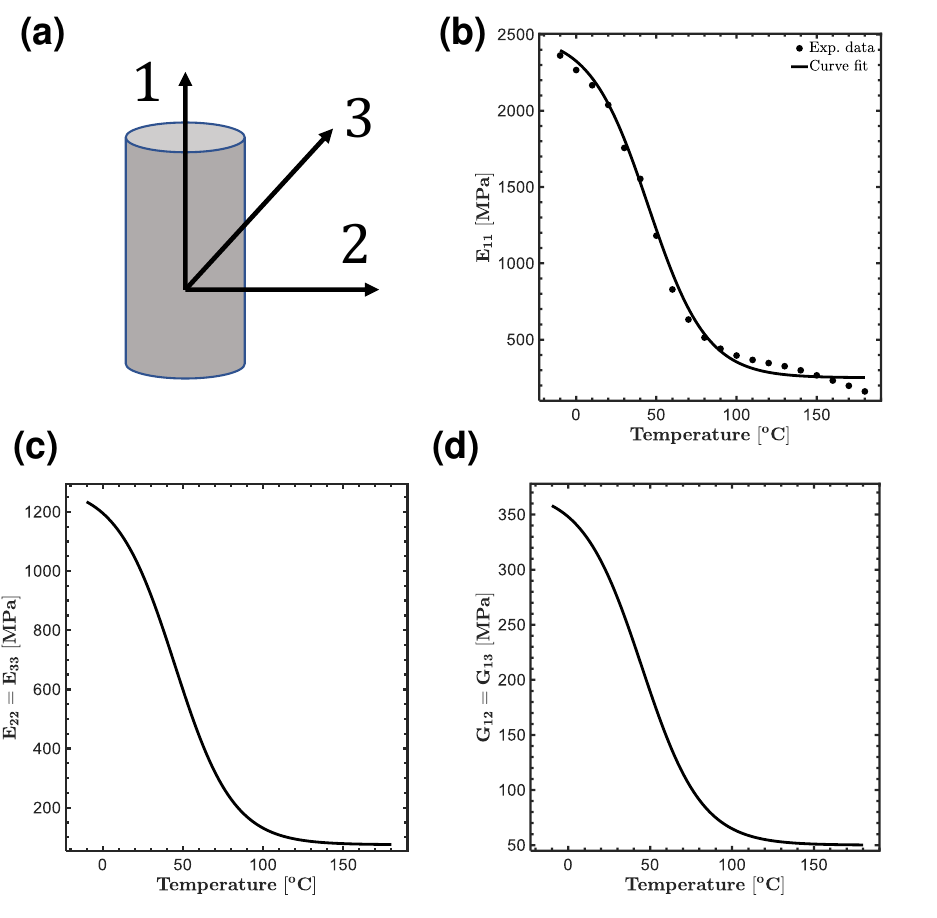}
    \caption{(a) Coordinate system for the precursor fiber, (b) the modulus along the fiber-axis, (c) elastic modulus in the radial or transverse direction, and (d) the shear modulus along the torsional direction of the fiber}
    \label{fig:TCPAMatPropsLiterature}
\end{figure}

The precursor fiber is annealed before it is used for the fabrication of the TCPA. The annealed fiber is tested along the axial direction under temperature sweep in a dynamic mechanical analyzer (DMA). Figure \ref{fig:TCPAMatPropsLiterature}(a) shows the coordinate system used to label the material orientation, Figure \ref{fig:TCPAMatPropsLiterature}(b) shows the temperature-dependent elastic modulus of along the fiber axis. A sigmoid function of the following nature is fit in this data for $E_{11}(T)$ as:
\begin{equation}\label{E(T)}
    E_{11}(T) = \left(E_{11}^g -E_{11}^r\right)\left(1-\xi(T) \right) +E_{11}^r 
\end{equation}

where, superscripts $g$ and $r$ denote the material property in glassy and rubbery regimes of the precursor fiber, respectively. The sigmoid function $\xi(T)$ is used to capture the glass-transition regime and is given as:
\begin{equation}\label{Sigmoid}
   \xi(T) = \frac{1}{\left(1+exp(-a(T-T_g)) \right)}
\end{equation}
The parameter $a$ in the sigmoid function dictates the width of the transition from glassy to rubbery state and  $T_g$ is glass transition temperature of the precursor fiber measured experimentally. The values of $E_{11}^g$, $E_{11}^r$ and $a$ is estimated by regression analysis of Eqn. \ref{E(T)} with the experimental data. The parameters obtained are shown in Table \ref{tab:Mat-props-Nylon66}. If we measure the modulus in other material directions under glassy and rubbery phases, it is reasonable to assume the same temperature dependence of those components as the $E_{11}$ with the sigmoid parameters estimated above. For instance, we obtained the value of transverse direction modulus $E_{22}=E_{33}$ in glassy and rubery phases from the work of Swartz et al. \cite{swartz2018experimental} and takes the following form and shown in Figure \ref{fig:TCPAMatPropsLiterature}(c): 
\begin{equation}\label{E22(T)}
    E_{22}(T) = E_{33}(T)= \left(E_{22}^g -E_{22}^r\right)\left(1-\xi(T) \right) +E_{22}^r 
\end{equation}
Similarly, we have measured the shear modulus of the annealed fiber along its polar direction  ($G_{12}=G_{13}$) in glassy and rubbery regimes, and we can use the sigmoid variation (Eqn. \ref{Sigmoid}) to estimate the temperature dependent shear modulus as (plot shown in Figure \ref{fig:TCPAMatPropsLiterature}(d)): 
\begin{equation}\label{G(T)}
    G_{12}(T) = G_{13}= \left(G_{12}^g -G_{12}^r\right)\left(1-\xi(T) \right) +G_{12}^r 
\end{equation}

The Poisson's ratio in all three directions is assumed to be temperature-independent $\nu_{12}=\nu_{13}=\nu_{23}=0.40$\cite{callister2020fundamentals}. The shear modulus ($G_{23}$) in the plane of isotropy for a transversely isotropic fiber can be evaluated using the Young's modulus ($E_{22}$) and Poisson's ratio ($\nu_{23}$). There are 5 independent engineering constants required to specify a transversely isotropic fiber (refer Appendix). In summary, the elasticity constants for the precursor fiber properties in local fiber coordinate systems are:
\begin{align}
    E_{11} &= E_{11}(T)\\
    E_{22}&=E_{33}=E_{22}(T)\\
    G_{12}&=G_{13}=G_{12}(T)\\
    \nu_{12}&=\nu_{13}=\nu_{23}=0.40\\
    G_{23}(T)&=\frac{E_{22}(T)}{2(1+\nu_{23})}
\end{align}

\begin{table}[h]
    \centering
    \caption{Coefficients of polynomial fits the elastic constants for Nylon 6,6 precursor fiber used in the model}
    \begin{tabular}{|c|c|c|c|c|c|c|c|c|}
    \hline
       $E_{11}^g$  & $E_{11}^r$ & $E_{22}^g$  & $E_{22}^r$ &$G_{12}^g$ &$G_{12}^r$ & $\nu_{12}=\nu_{13}=\nu_{23}$& $T_g$ & $a$ \\
       \hline
       $2390$ MPa & $250$ MPa & $1290$ MPa & $75$ MPa & $373$ MPa & $50$ MPa & $0.40$ &  $45$ $^o$C & 0.1 $^o$C$^{-1}$\\
       \hline
    \end{tabular}
    \label{tab:Mat-props-Nylon66}
\end{table}

\begin{figure}[h]
    \centering
    \includegraphics[trim={0.cm 10.5cm 16.8cm 0.8cm},clip=true ,width=0.980\textwidth]{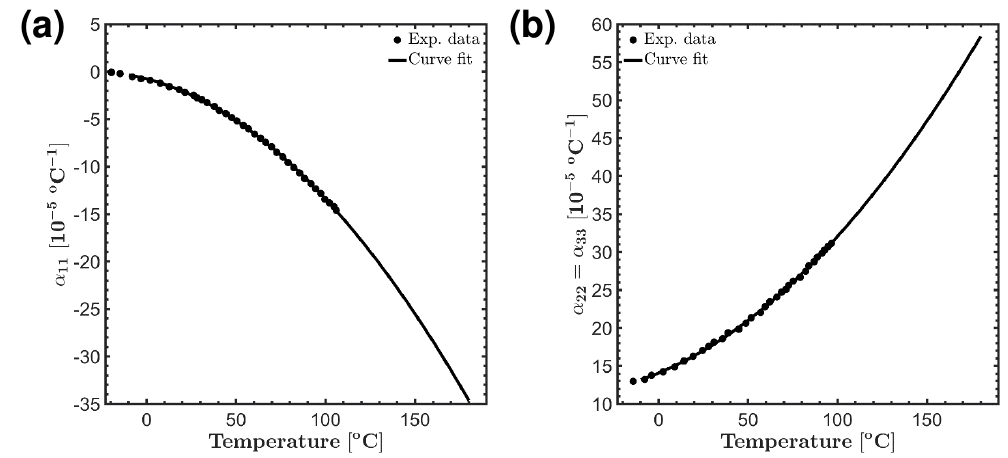}
    \caption{(a) The coefficient of thermal expansion  along the fiber axis ($\alpha_{11}(T)$) and (b) in transverse or radial directions ($\alpha_{22}(T)=\alpha_{33}(T)$), the data points are considered from Choy et. al\cite{Choy1979,Choy1981}}
    \label{fig:TCPAMatPropsAlphaLiterature}
\end{figure}

 The thermal expansion coefficients of the highly drawn Nylon 6,6 fiber are shown in Figure \ref{fig:TCPAMatPropsAlphaLiterature}(a)-(b), where the data points are taken from the work of Choy et al., and a second order polynomial is fit and extrapolated to higher temperature ranges\cite{Choy1979, Choy1981}. The axial ($\alpha_{11}(T)$) and radial ($\alpha_{22}(T)=\alpha_{33}(T)$) CTE take the following form:
 \begin{align}\label{axial-radial-CTE}
     \alpha_{11}(T)&=a_2T^2+a_1T+a_0\\
     \alpha_{22}(T)=\alpha_{33}(T)&=b_2T^2+b_1T+b_0
 \end{align}

 where, the coefficients $a_i$ and $b_i$ are obtained by fitting the curve shown in Figure \ref{fig:TCPAMatPropsAlphaLiterature}(a) and (b), respectively, and given in Table \ref{tab:CTE-Nylon66}. 
\begin{table}[h]
    \centering
    \caption{Coefficients of second-order polynomial fit for axial and radial thermal expansion coefficient for Nylon 6,6 used in the model}
    \begin{tabular}{|c|c|c|c|c|c|}
    \hline
       $a_2 ~(10^-4~ ^oC^{-3})$  & $a_1 ~(10^-2~ ^oC^{-3})$ & $a_0~ (^oC^{-1})$  & $b_2~ (10^-4~ ^oC^{-3})$  & $b_1 ~(10^-2 ~^oC^{-3})$ & $b_0 ~(^oC^{-1})$ \\
       \hline
       $-7.711$  & $-4.941$ & $-0.7591$  & $8.288$ & $9.698$ &$14.08$ \\
       \hline
    \end{tabular}
    \label{tab:CTE-Nylon66}
\end{table}
 
\subsection{TCPA geometry and choice of mesh}
\begin{figure}[h]
    \centering
    \includegraphics[trim={0.0cm 2.0cm 5.5cm 0.6cm},clip=true ,width=0.9\textwidth]{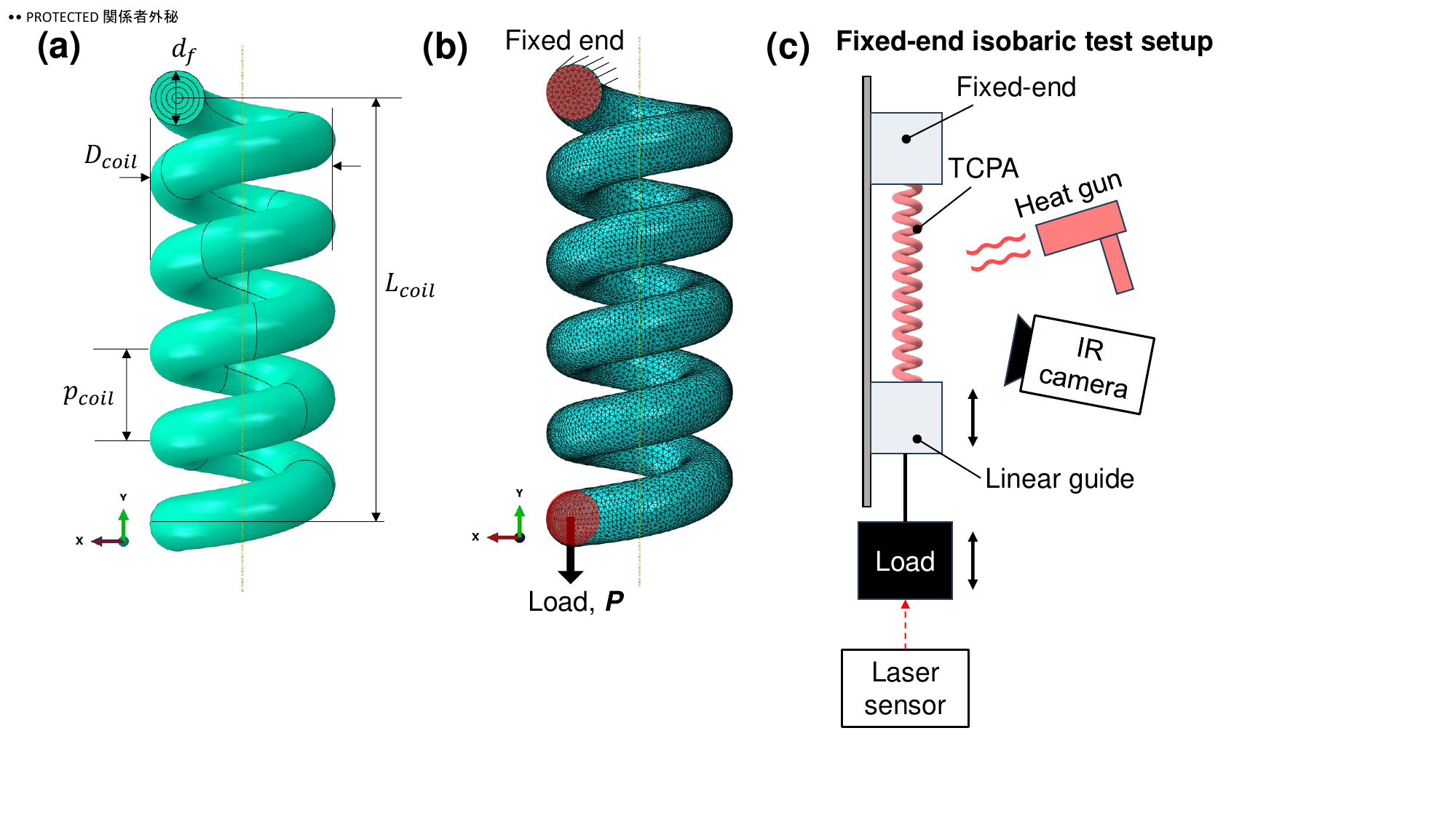}
    \caption{(a) Geometry of a TCPA coil with concentric helical annular partitions showing different coil dimensions, (b) the finite element mesh showing the boundary conditions at the top and bottom ends of the coil and (c) schematic of the experimental setup for testing TCPA under isobaric loading condition}
    \label{fig:TCPAFEandGeometry}
\end{figure}

The coil is created by revolving a circular cross-section about the coil axis and is partitioned into 5 concentric helical annulus and its adequacy is checked in the results section. Figure \ref{fig:TCPAFEandGeometry}(a) shows the  geometry of 5 coils modeled. The TCPA coil geometry is defined by the fiber diameter ($d_f$), coil diameter ($D_{coil}$), and coil pitch ($p_{coil}$), therefore, the total length of the coil considered in the FE model is $L_{coil} = 5\times p_{coil}$. The top face shows five partitions created and their respective orientations are assigned, the procedure for the same is described in the next subsection. The 3D geometry is discretized using second-order tetrahedron elements (C3D10 of ABAQUS\cite{smith2009abaqus}) mesh (shown in Figure \ref{fig:TCPAFEandGeometry}(b)). The nodes on the top face are constrained in all translation directions ($u_x=u_y=u_z=0$) i.e. clamped. The bottom face nodes are tied to a rigid-body reference node which is constrained in rotation about global $y-axis$ and a load is applied downward as performed experimentally.

\subsection{Local material orientation assignment}

As described in the previous subsection, the TCPA can be modeled as a set of the finite number of concentric laminate made up of the same ply with a rotated material orientation by an angle described by Eqn. \ref{alpha_eq}. Figure \ref{fig:twistSchematic}(d) shows a coil with concentric partitions. Now, the objective is to assign a local material orientation to material points of the TCPA based on their radial location from the helix axis. It can be achieved by writing a material orientation subroutine but ABAQUS/CAE offers a simpler approach. We utilized the simplicity of the helix geometry to assign the material orientation in this problem. We used the discrete material orientation option in ABAQUS/CAE\cite{smith2009abaqus} where we use the normal of the coil surface as one direction ($3-axis$), the helix axis as the second direction ($2-axis$), then the third direction ($1-axis$) is automatically determined that aligns along the circumference of the helix at a fixed location on its axis (shown in Figure \ref{fig:twistSchematic}(d)). Then the twist angle is assigned by rotating this coordinate system about $3-axis$ (local radial direction) which is normal to the coil surface. The solid arrows show the orientation computed based on geometry and the dashed arrows indicate the rotated direction after assigning the angle of twist about $axis-3$. This rotation angle ($\alpha(r)$) is computed for a partition based on its mean radial location and shown in Figure \ref{fig:twistSchematic}(d).

\subsection{Fabrication and Testing of TCPA}

The twisted and coiled polymer actuator sample was fabricated by twisting and coiling a precursor fiber. The fiber used in this study is a 1 mm diameter nylon fiber (from *Shaddock Fishing*, clear monofilament fishing line, 57.47 kg test strength) with a density of 1 g/cm$^3$. The fiber is first twisted until the instability is initiated. The fiber starts snarling while a 1 kg load is applied at one end to prevent undesired local writhing. It is then coiled around a metal pin with a diameter of 3.35 mm to form a uniform helix. The twisted and coiled sample is subsequently annealed in oil at 120°C for two hours to partially release the residual stress\cite{WANG2024109440}. Following this, the sample is untwisted at one end and stretched until the pitch between the coils reaches 4.57 mm. Finally, the actuator is annealed again in oil at 170 $^o$C with both ends fixed for six hours, which permanently fixed the shape of the sample.
The fixed-end isobaric test schematic is shown in Figure \ref{fig:TCPAFEandGeometry}(c). We attach the TCPA to the linear guide to prevent end rotation. One end of the muscle is fixed while the other, connected to a load, is free to move in the vertical direction. The muscle is heated up by a heat gun for temperature uniformity, and the muscle temperature is measured by an IR camera. The displacement of the load is measured by the laser sensor.  

\section{Results and Discussion}

The developed FE model is checked for basic features of the TCPA actuators. Validation of the FE model is necessary and helps to ensure whether the model is able to capture the intended physics of the actuator. The coil geometry and the temperature-dependent material properties, including in the cylindrical laminates, make the problem highly non-linear. In this section, we 
 start with model validation against experiments and subsequently present the results and their implications for actuation mechanisms from various parametric studies.

\subsection{Model Validation}
The continuum scale model of the TCPA is implemented using the finite element framework. The temperature-dependent material properties of the precursor fiber as outlined in Section \ref{mat-prop-section}, are used for all simulations in this work, unless specified. The coil parameters considered for validation study are shown in Table \ref{tab:Val_coil}, and the meaning of the coil parameters can be referred from Figure \ref{fig:TCPAFEandGeometry}. A small load of value $P=0.1~N$ (~10 g) is applied on one end of the coil and the other end is kept fixed in all directions. The loaded end is constrained in rotation about the axis of the coil to accommodate the linear guide in the experimental setup. Initially, the coil is kept at $23~^o$C and it is loaded at this temperature by applying the load $P$. After the coil deforms under static loading conditions, the temperature is raised to a target value of $122~^o$C under static analysis in ABAQUS/Standard. In this case, the actuation stroke is measured as the percent change in length with respect to the loaded state of the TCPA coil.
\begin{table}[h]
    \centering
    \caption{Coil parameters used for validation of the model}
    \begin{tabular}{|c|c|c|c|}
    \hline
       $D_{coil}$  & $d_f$ & $p_{coil}$ & $\alpha_f$ \\
       \hline
       $4.35$ mm & $1$ mm & $4.57$ mm & $31^o$ \\ 
       \hline
    \end{tabular}
    \label{tab:Val_coil}
\end{table}
\begin{figure}[ht]
    \centering
    \includegraphics[trim={0.cm 0.0cm 15.0cm 0.5cm},clip=true ,width=0.6\textwidth]{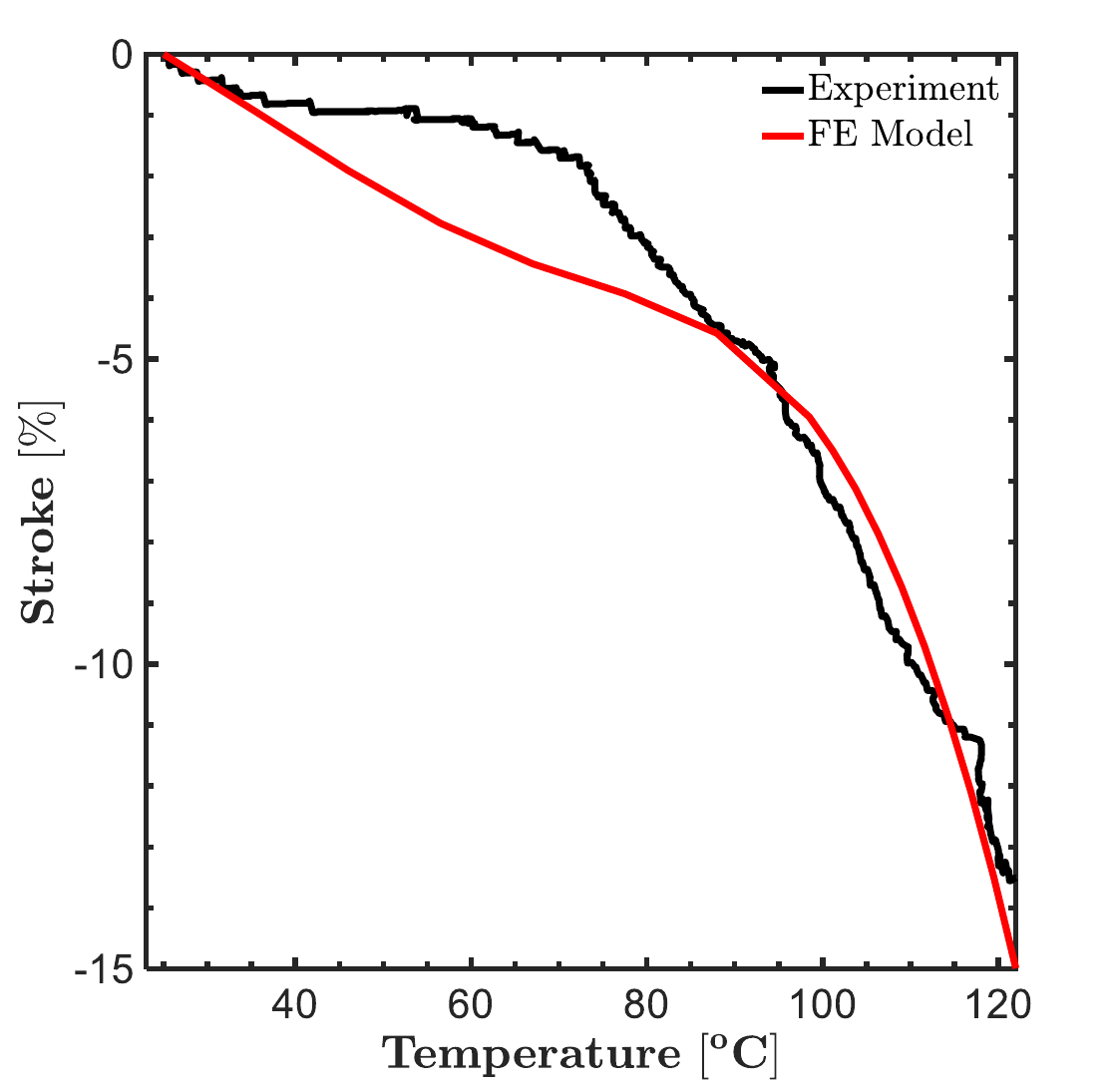}
    \caption{The actuation stroke of the TCPA muscle vs. temperature from FE model and its comparison with experimental values under the application of a 10g weight}
    \label{fig:FE_vs_Exp_TCPA}
\end{figure}

Figure \ref{fig:FE_vs_Exp_TCPA} shows the comparison of the model actuation with the values measured experimentally as the TCPA is heated. The FE model is able to capture the actuation response of the coil and is in close agreement with the experimentally observed response. Since this is a highly non-linear problem to model as it includes the creeping of the TCPA as well which we do not model in this work. However, the model captures the initial temperature-driven softening of the TCPA up to $80~^o$C, even though experimentally, it shows more softening which might be due to the creep deformation of the muscle. Overall, the agreement of the model with experiments gives us the confidence to look into parametric studies to understand the actuation mechanisms in further detail. 

\subsection{Chirality of TCPA}

\begin{figure}[h]
    \centering
    \includegraphics[trim={0cm 0.0cm 2.25cm 1.0cm},clip=true ,width=0.98\textwidth]{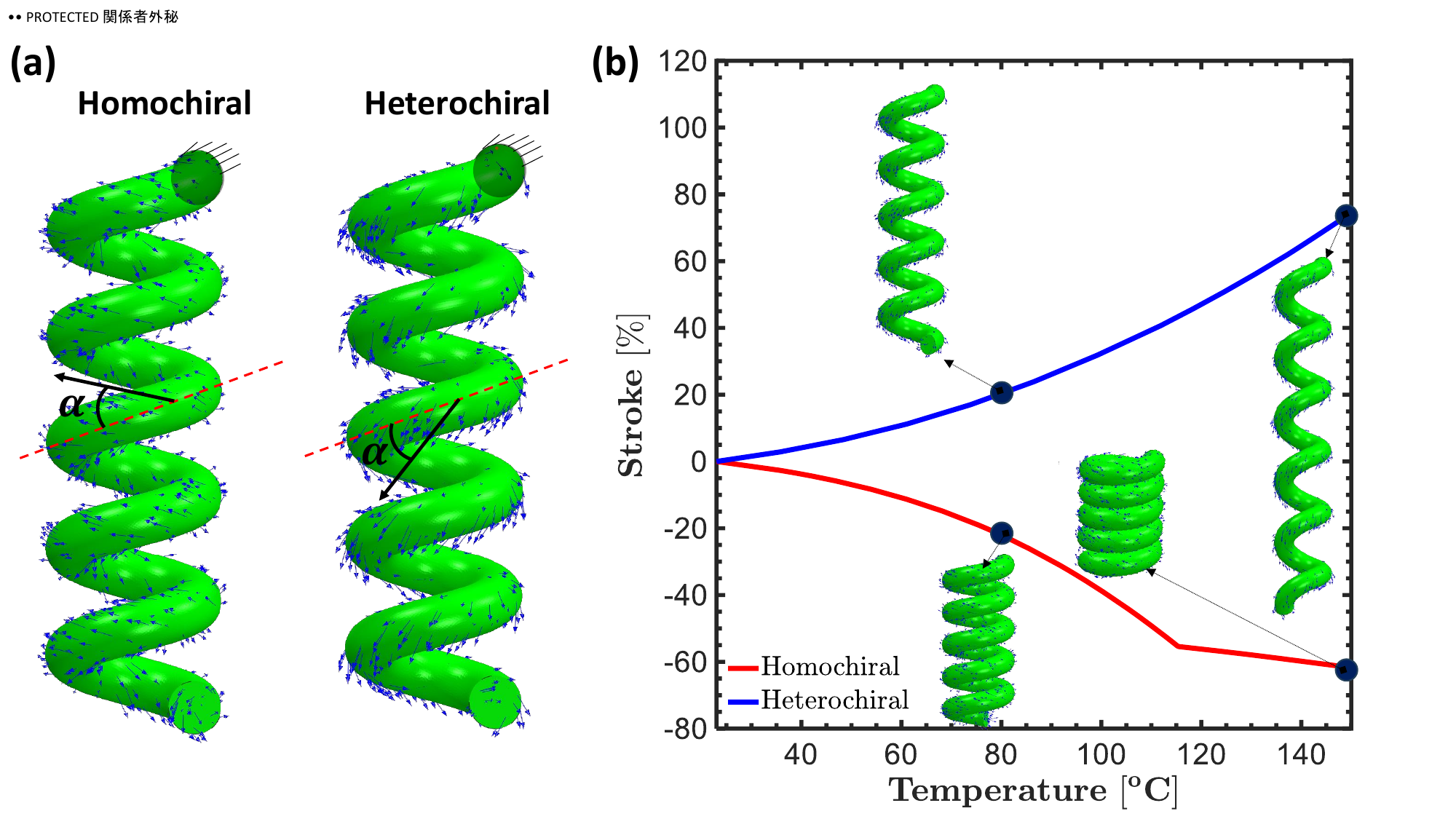}
    \caption{(a) The crystalline axis (blue arrows) direction w.r.t coil leads to chirality of the muscle, (b) free actuation response for homochiral (contraction) and heterochiral (extension) using the FE model; insets show the deformed states of the two types of coils}
    \label{fig:TCPAchirality}
\end{figure}

To ensure the adequacy of the FE model, it can be tested for some special twist angles of the muscle that defines the nature of stroke. There are two ways to introduce the twist in the fiber during the fabrication process: homochiral and heterochiral. When the twist in the muscle is introduced in the same direction as the orientation of the helical coil, this is referred to as homochiral muscle (shown in Figure \ref{fig:TCPAchirality}(a))\cite{haines2014artificial}. However, when it is inserted opposite to the coil-helix, it is called heterochiral muscle\cite{haines2014artificial}. The homochiral configuration results in a contracted actuation while the heterochiral muscle shows extension upon heating\cite{haines2014artificial,haines2016new}. 

\begin{table}[h]
    \centering
    \caption{Coil parameters used for parametric studies}
    \begin{tabular}{|c|c|c|c|}
    \hline
       $D_{coil}$  & $d_f$ & $p_{coil}$ & $\alpha_f$ \\
       \hline
       $3.36$ mm & $1$ mm & $2$ mm & $42^o$ \\ 
       \hline
    \end{tabular}
    \label{tab:Val_coil}
\end{table}

Note the direction of the blue arrows that indicate the reoriented fiber axis material direction of the twist on the surface of the coil (rotated $1-axis$ in Figure \ref{fig:twistSchematic}(d)). Figure \ref{fig:TCPAchirality} (b) shows that the developed FE model captures the chirality dependence of the introduced twist with respect to the coil-helix. As the TCPA actuates upon heating, the crystalline axis (blue arrow) gets aligned with the coil helix thereby the coil shifts upwards closing the gaps between coils for homochiral case. This is Due to the mismatch in thermal coefficients between concentric laminates, equilibrium states needs to have a reoriented material direction to minimize the stresses in the heated configuration. On the other hand, for the case of heterochiral muscle, the coils increase the separation between them to achieve a state of oriented material directions (note the blue arrows align nearly vertical at 140 $^o$C). This can be further understood by a simpler geometry, a straight precursor fiber  that has a large twist. It can be thought of as concentric laminates with radially varying material orientations. Upon heating such a fiber, the difference in thermal expansion in axial and radial direction will result in a coupling between its axial and torsional response\cite{higueras2020finite}. Therefore, as the fiber will contract axial upon heating, there will be a torsional twist about its axis that means the fiber would have a tendency to untwist from its originally twisted configuration in order to minimize the interfacial shear stresses. Now a coil made out of a twisted fiber will result in a coupling between the axial, torsional, and bending of the fiber axis due to its curved geometry. Therefore, the contraction of the coil is a bulk motion as a result of the bending of the coil axis. However, the direction of bending depends on the twist direction about the coil axis and that is how we define the chirality of the TCPA muscle. For a heterochiral TCPA, the bending occurs in the direction that results in the separation of two coils from each other, therefore, an extension of the muscle upon heating. 

\subsection{Influence of number of partitions}
\begin{figure}[ht]
    \centering
    \includegraphics[trim={0.cm 1.5cm 2.8cm 0.6cm},clip=true ,width=0.95\textwidth]{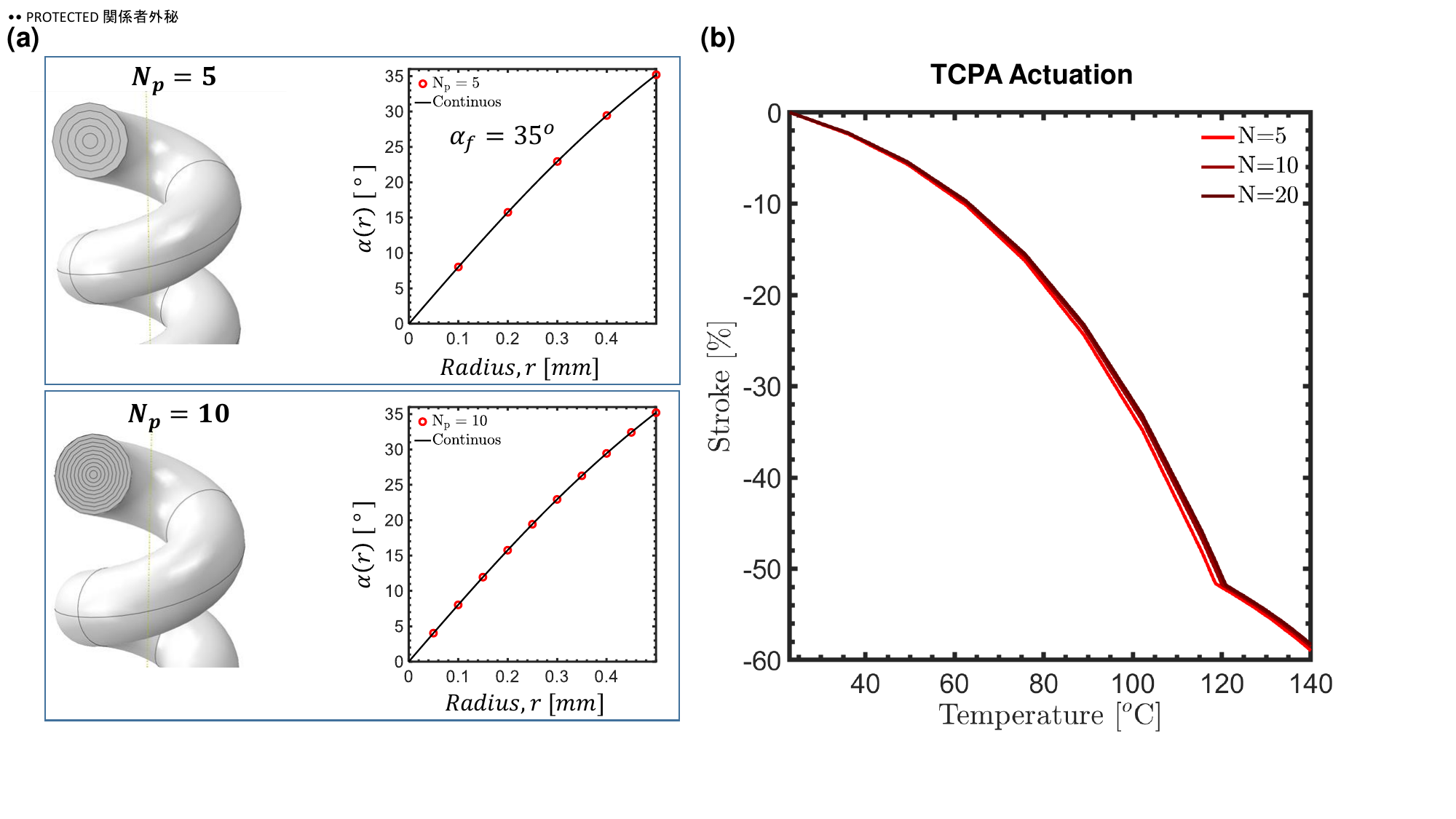}
    \caption{(a) Two cases of coil geometry showing the number of concentric partitions, and their respective distribution of  orientation as a function of fiber radius and (b) shows the contraction of the TCPA upon heating measured w.r.t its undeformed length}
    \label{fig:NPartitions}
\end{figure}

A key assumption made in this FE model is that the TCPA can be represented as a finite number of concentric helices. Therefore, we tested the TCPA response for the cases when we divide the cross-section into 5, 10 or 20 concentric partitions (shown in Figure \ref{fig:NPartitions}(a)). The discretized twist angle for each annulus is computed at its mean radius. Figure \ref{fig:NPartitions}(b) shows the contraction as a percent of the coil's initial length. It is evident that the 5 partitions are enough to represent the concentric helices FE model. The lower number of partitions case shows more contraction which is expected as thicker annuli have larger twist angles assigned to them since the twist angle value is computed at the outer radius of the annulus. This difference grows more with high temperature as the larger deformation occurs at a higher temperature, therefore, a more pronounced difference in the actuation response between the two cases. However, the difference is less than $5\%$ at the maximum temperature location which is reasonable as the lesser number of partitions model saves on FE mesh size.

\subsection{Understanding the actuation of TCPAs}

\subsubsection{Influece of twist angle}

\begin{figure}[h]
    \centering
    \includegraphics[trim={0.cm 3.5cm 1.8cm 0.6cm},clip=true ,width=0.98\textwidth]{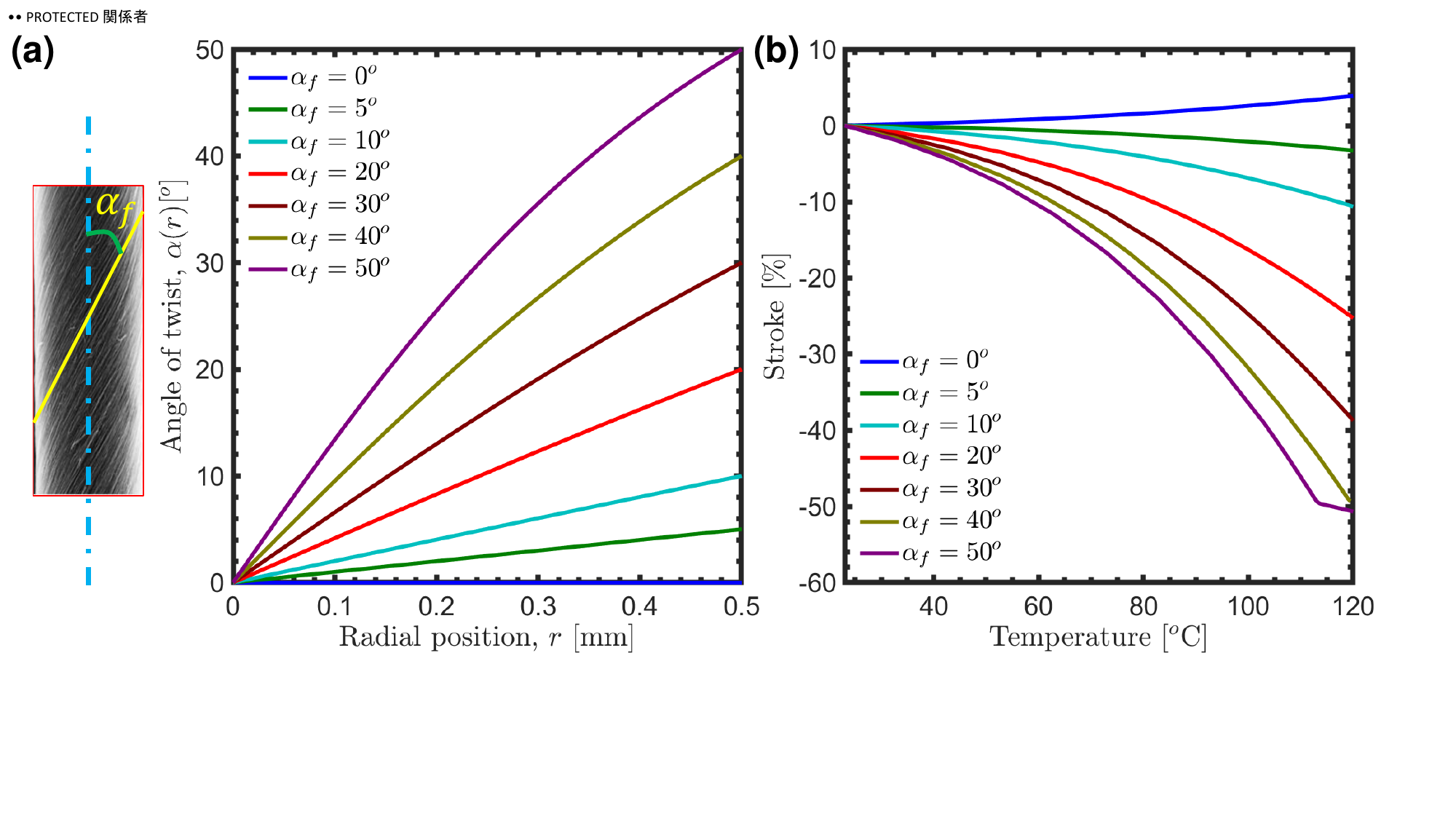}
    \caption{(a)The value of angle of twist a function of radius in the concentric laminate model and (b) the corresponding free actuation response of the TCPA muscle}
    \label{fig:TCPATwistAngle}
\end{figure}

An important feature the FE model should capture is the extent of the twist on the fiber. Figure \ref{fig:TCPATwistAngle} shows that the free actuation response for a homochiral muscle increases as more twist is introduced in it. However, this advantage will start stagnating as we approach more towards $\alpha_f=90^o$ which is not practical to fabricate from a Nylon fiber. Furthermore, the precursor fiber needs to undergo very large permanent deformation in torsional shear mode to attain a high value of twist angle. The high value of twist is introduced such that the precursor fiber doesn't fracture. The case of $\alpha_f=0$ shows that the muscle is not able to contract, even though small, but, it extends upon heating. This behavior is in line with the experimentally observed effect of angle of twist in literature\cite{haines2014artificial,Saharan2019} and in other computational models\cite{Yang2016TCPA,Hunt2021}. This is the due to the fact the coiling fiber itself introduces a small yet finite twist between its two ends and even with $\alpha_f=0$ in the precursor fiber makes it behave like a heterochiral muscle. 

\subsubsection{Role of material anisotropy}

\begin{figure}[h]
    \centering
    \includegraphics[trim={0.5cm 5.8cm 3.cm 0.8cm},clip=true ,width=0.98\textwidth]{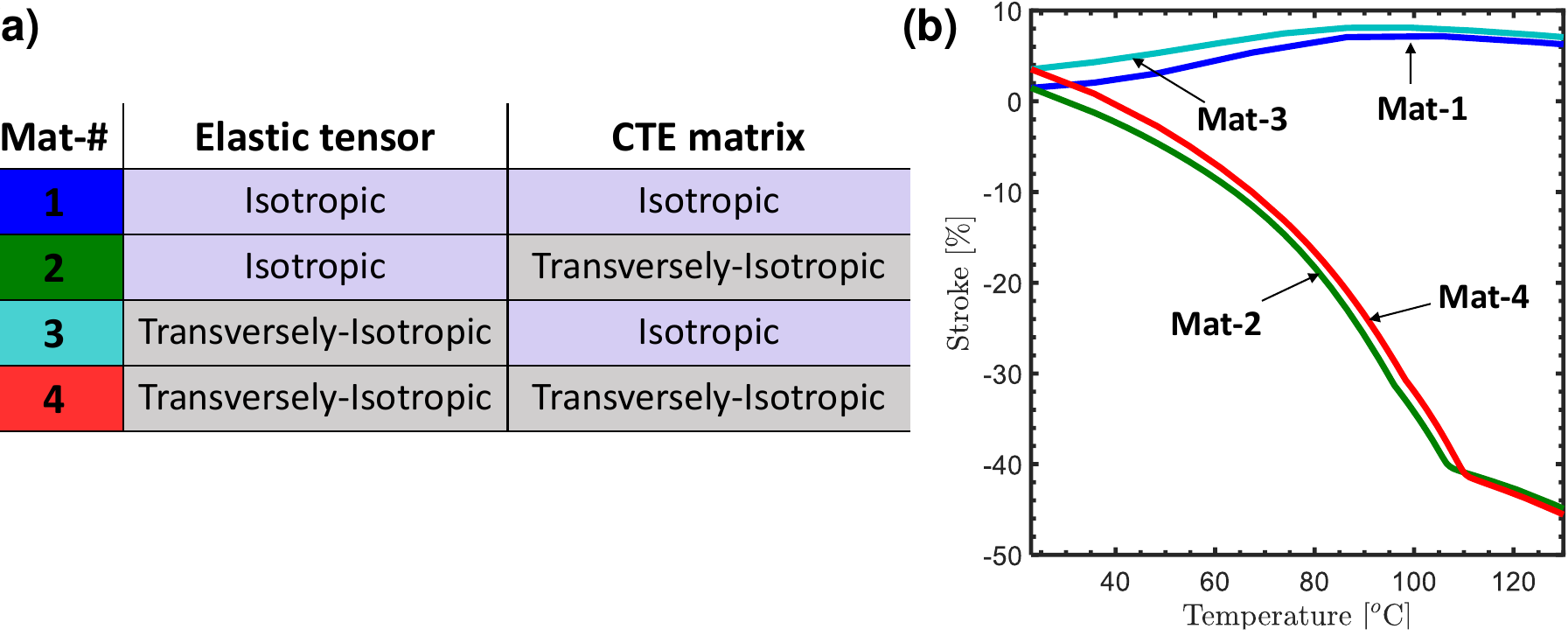}
    \caption{TCPA muscle material  with anisotropy in elastic tensor or thermal expansion (CTE) matrix: (a) material combinations considered, and  (b) their actuation response}
    \label{fig:TCPAMatAniso}
\end{figure}

This has been hypothesized that the origin of the large contraction in the anisotropy precursor fiber. In this work, we look at the role of material anisotropy systematically and make an attempt to understand what are the key parameters responsible for the large actuation of the TCPA actuators. As we have shown in previous sections that the FE model captures the physics of the actuation and adequate for further investigations. Therefore, we can consider various material parameter combinations that may not be viable to fabricate experimentally. There are two sources of material anisotropy: one in the elasticity tensor, and the second lies in the coefficient of thermal expansion matrix. Therefore, there are primarily four combinations, where we assume isotropy in one of the properties. Figure \ref{fig:TCPAMatAniso}(a) shows the four material cases for which we simulate the TCPA actuation response. The baseline material is Mat-4 where both elasticity and CTE matrix has anisotropy and the properties are considered that of the fiber in previous sections. The isotropic elasticity case (Mat-1 and Mat-2) is assumed to have the modulus along the fiber direction (Figure \ref{fig:TCPAMatPropsLiterature} (b)) with Poisson's ratio of 0.4. The fiber axis CTE is assumed for the case of the isotropic CTE matrix i.e. for Mat-1 and Mat-3 (plotted in Figure \ref{fig:TCPAMatPropsAlphaLiterature}(a)). Both elastic tensor and CTE are assumed to have temperature dependence as in precursor fiber properties.

Figure \ref{fig:TCPAMatAniso}(b) shows the actuation response for the four cases of material anisotropy in the fiber with 10 g of load. The material case Mat-1 where both properties are isotropic is basically the case of a TCPA muscle fabricated using an isotropic material with negative CTE. Such a fiber does not show any contraction, instead, it extends upon heating as a result of the elastic modulus softening. Beyond 80 $^o$C, the muscle shows small contraction that is because of the overall volumetric shrinkage of the fiber as the isotropic CTE assumed is negative for this material case. The case of anisotropy in CTE matrix shows the contraction despite the isotropy in the elasticity. The Mat-2 and Mat-4 show a very close actuation response with a minimal difference owing to the difference in elastic constants between the two cases. The Mat-2 shows a smaller pre-stretch in the initial loaded state whereas Mat-4 shows larger static deformation at room temperature (23 $^o$C). This is due to the fact that the isotropic elasticity tensor shows higher stiffness along the fiber axis as the angle of twist doesn't reduce the effective axial stiffness with isotropic elasticity tensor in case of Mat-2. Additionally, comparing the total stroke, Mat-2 shows a slightly shorter stroke as compared to Mat-4 and this is due to its higher elastic stiffness and is discussed further in next section. Comparing Mat-3, despite having a negative CTE although isotropic, shows just extension upon heating. It shows more extension as compared to Mat-1 which can be attributed to it's lower elastic stiffness as compared to Mat-1. These four material cases provide an insight into the key features a precursor fiber should have in order to develop a TCPA muscle that can produce a large actuation. 
Overall, the anisotropy in CTE is responsible for the actuation irrespective of the type of material symmetry in the elasticity tensor of the precursor fiber.

\subsubsection{Influence of extent of material properties}
\begin{figure}[h]
    \centering
    \includegraphics[trim={0.cm 0cm 5.8cm 0.0cm},clip=true ,width=0.95\textwidth]{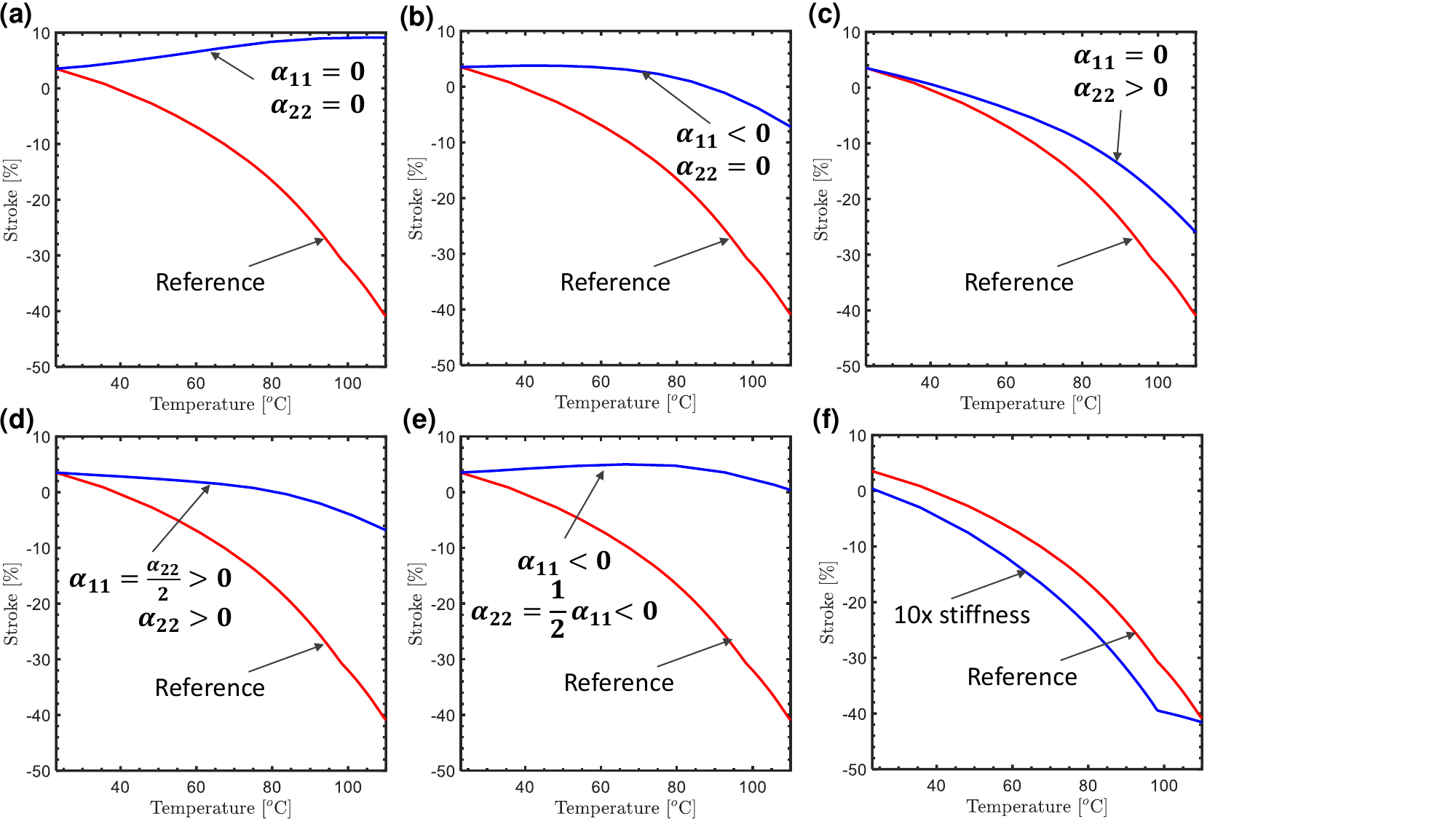}
    \caption{Effect of material properties and comparison against the reference fiber material: (a) axial and radial direction CTEs are 0, (b) axial as in reference i.e. negative but radial is 0, (c) axial is 0 but radial as in reference, (d) radial as in reference but axial is positive, (e) axial is reference but radial is negative, (f) increase the stiffness tensor by 10 times of reference}
    \label{fig:TCPAMatCTEeffct}
\end{figure}

In the previous section, we discussed the role of material anisotropy towards actuation and observed that the anisotropy in CTE is the primary factor in actuation of TCPAs.  In this section, we probe the extent of thermomechanical material properties on the actuation response of a TCPA. In the previous sections, we considered temperature dependent CTE values of the precursor fiber, where axial CTE was negative while the radial direction CTE was positive. Figure \ref{fig:TCPAMatCTEeffct}(a) shows the case of a non-expandable material i.e. the CTE in all directions is assumed to be zero. The model shows the muscle just extends upon heating which is attributed solely to the softening of the elasticity tensor. The reference material case has the properties of a Nylon 6,6 as outlined in method section. Next, we consider a case where we keep the axial CTE to be the same as reference material, however, the radial direction CTE is assumed to be zero (shown in Figure \ref{fig:TCPAMatCTEeffct}(b)). This TCPA shows actuation even though smaller as compared to the reference material. This demonstrates that the radial swelling is not the only factor responsible for actuation in TCPAs. Similarly, we consider a case with non-expanding material along the fiber axial direction but radial expansion is kept as that of a reference material.

Figure \ref{fig:TCPAMatCTEeffct}(c) shows that the actuation has improved as compared to non radial expansion case. This points outs out that the axial contraction is not the key factor responsible for the large actuation as hypothesized by several previous studies\cite{karami2017modeling,haines2014artificial,zhang2017modeling}. This is further proved by considering a case where the axial expansion is positive but half in magnitude of the reference radial expansion values (Figure \ref{fig:TCPAMatCTEeffct}(d)). Such material can be thought of as a matrix with non-expanding short fibers embedded in it with axial orientation. Such fiber can also show actuation yet smaller. Another case is considered with a hypothetical material with negative axial expansion and also a negative radial but half in magnitude of the reference axial CTE. This case demonstrates a very small extent of  actuation as shown in Figure \ref{fig:TCPAMatCTEeffct}(e). This indicates the TCPAs can be fabricated with volumetrically contracting material as well but the radial direction CTE needs to be lower in magnitude than axial direction for a homochiral and higher in magnitude for heterochiral muscle. The results from these cases indicate that the extent of mismatch in the CTE values in radial and axial direction i.e. $(\alpha_{11}-\alpha_{22})$ matters the most in terms of the actuation despite the contacting or expanding nature of thermal expansion. Figure \ref{fig:TCPAMatCTEeffct}(f) shows a case with a assumed a material case high stiffness (10 times the elasticity tensor that of Nylon 6,6) while the the CTE is kept the same as reference material. This shows that a stiffer material with the same CTE matrix does not alter the actuation response, however, it enhances the load carrying capability of a TCPA. If we compare the actuation with respect to loaded state (initial point in Stroke axis), the stiffer material case shows a lower contraction and this is due to the fact that it attains the coil-coil contact sooner as compared to the reference material. The stiffer material case coil is shorter in total length in the loaded state at room temperature before heating. This is in-line with experimentally demonstrated for different fibers by Haines et al.\cite{haines2014artificial}. This provides us a guidance into the choice of initial coil pitch for a stiffer material. If a material is softer, a lower coil pitch should be preferred, however, if the precursor fiber is made out of a stiffer material, a high pitch coil is desirable for improving both stroke and load carrying capability.
This parametric study provides the insights into the choice of precursor fibers in order to develop high performance TCPAs.


\subsubsection{Core-shell composite TCPA}
\begin{figure}[h]
    \centering
    \includegraphics[trim={0.25cm 6.5cm 0.60cm 0.8cm},clip=true ,width=0.98\textwidth]{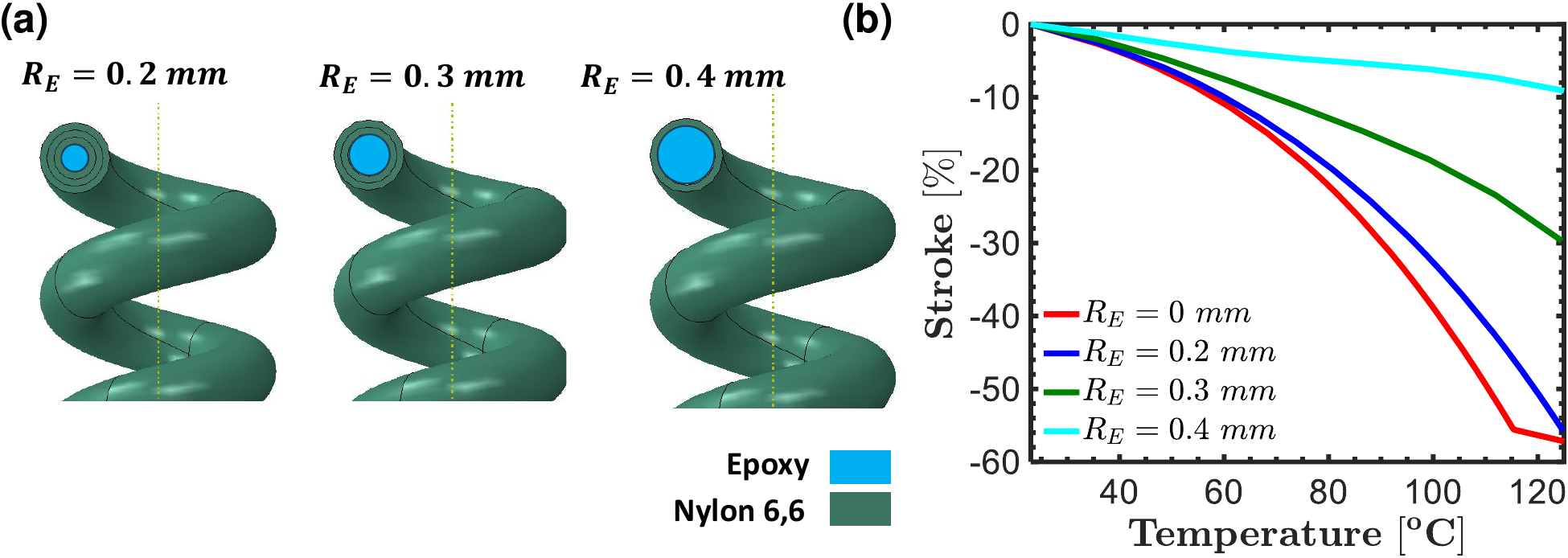}
    \caption{A core shell composite-based TCPA: (a) inner core of epoxy material with different radii, and (b) their actuation response. $R_E$ denotes the radius of the epoxy core inside a fiber of radius 0.5 mm.}
    \label{fig:FE_coreshell_TCPA}
\end{figure}

The actuation response of a TCPA, as discussed in previous sections, is derived from the difference in thermal expansion coefficient in the concentric laminate due to the twist and enhanced by the coil geometry. One of the major limitations of semi-crystalline polymer-based TCPAs (Nylon 6,6 or Polyethylene ) is the creep under higher loading and temperature regimes and/or under long duration of loading\cite{swartz2018experimental,WANG2024109440,tsai2025high,guo2025driving}. The creep is a phenomenon where the material deforms permanently over time upon the application of a constant stress. The creep behavior exacerbates at elevated temperatures and under the application of high stress which is a common loading case for these actuators. Epoxies are thermoset polymers that are known to show negligible creep response and are used for high-performance structural applications\cite{sa2011creep,singh2017damage,singh2024understanding,singh2023computational}. Based on the understanding of the actuation physics using the three-dimensional FE model, we propose a core shell composite TCPA where the center core is made of a creep-resistant material such as epoxy or its nanocompsite while the outer layers of the coil is twisted semi-crystalline polymer (shown in Figure \ref{fig:FE_coreshell_TCPA}(a)). The epoxies are less prone to creep due to their three-dimensional cross-linked network of polymer chains. The core material can be any material which is not susceptible to creep and it maintains its stiffness in the operational temperature range of a TCPA. Furthermore, this core material can be some conductive material that can act as a Joule heater for electro-actuation such as epoxy nanoscomposites.

Owing to its full geometric and material description capturing capabilities, our 3D FE model can be used to simulate such composites and analyze their performance as compared to the baseline TCPA based on Nylon 6,6. Figure \ref{fig:FE_coreshell_TCPA}(a) shows the various radii of the epoxy core inside a twisted and coiled Nylon 6,6 twisted and coiled hollow tubes of outer radius 0.5 mm. In this model, the interface between epoxy core and Nylon tube is assumed to be perfectly bonded. The Young's modulus, Poisson's ratio and coefficient of thermal expansion for epoxies are assumed to be 1 GPa, 0.35, and $3.5\times 10^{-5}$ $^o$C$^{-1}$, respectively. The free actuation response of these composite TCPA (as shown in Figure \ref{fig:FE_coreshell_TCPA}(b)) is comparable to pure Nylon-based TCPAs for lower value of core radius, and it starts reducing as we increase the isotropic epoxy core area fraction in the TCPA coil. Additionally, as we have seen with increasing stiffness of the coil material, increasing the coil pitch leads to higher actuation stroke. Another choice of core could be epoxy soaked carbon fiber-based TCPA\cite{lamuta2018theory} where it is wrapped in an outer Nylon-based hollow twisted tubes in a coiled configuration. Lamuta et al. developed a carbon fiber-based TCPAs and the present model can be used to investigate such actuators to improve their performance. This study provides us an alternative composite TCPA to address the issue of creep and load carrying capability of the Nylon-based actuators. There is a trade off between increasing the creep resistance and the amount of stroke. This work provides a framework to explore this vast design space for developing  high-performance TCPAs that demonstrate large stroke, creep resistant, and high load carrying capability. Furthermore, this analysis can be extended to other kinds of twisted and coiled polymer actuators to understand their mechanics and enhance their performance such as carbon nanotubes-based\cite{lee2017electrochemically}, carbon fiber-based\cite{lamuta2018theory}, hydrogel-based\cite{sim2020self,zhang2021reversible}TCPAs.


\section{Conclusions}
In this work, we conducted computational studies of the semi-crystalline polymer-based twisted and coiled polymer actuators (TCPA). A finite element model that accounts for the fabrication process, is developed for the TCPA and used to understand the fundamental factors responsible for the large actuation stroke of these actuators. The following conclusions can be drawn from this study:
\begin{itemize}
    \item  A concentric laminate model is implemented  for twisted and coiled semi-crystalline polymer-based TCPAs. The continuum-scale model captures the three-dimensional geometric and temperature dependent thermomechanical transversely-isotropic material description of radially varying twist inserted in the precursor fiber during the fabrication of the actuators, and is simulated in a finite element method solver. 
    \item The model is validated against actuation response experiments of fabricated TCPAs. The results show that model is in good agreement with  experiments and we demonstrate that it captures the physics of the actuators by considering the key aspects of the TCPAs such as the influence of chirality of the coil and angle of twist.
    \item The model is used to investigate the key factors responsible for the actuation of the TCPA muscles. A systematic parametric study show that the anisotropy in the CTE matrix plays a major role in TCPAs' actuation irrespetive of the anisotropy in the elasticity tensor of the precursor fiber. 
    \item Probing into the role of extent of material anisotropy reveals that absolute value of mismatch in  CTE matrix components plays a significant role irrespective of the contacting or expanding nature of the thermal coefficient in fiber axis or radial directions of the TCPAs.
    \item A creep-resistant core-shell composite TCPA is proposed and analyzed using the simulations for its performance as compared to conventional semi-crystalline polymer-based TCPAs. The results show that there is a trade off of stroke and creep resistance behavior, however, this model can be used to explore the design space to optimize their large stroke and high load carrying capability of TCPAs while mitigating the creep response.
    \item The framework developed in this work opens the scope to investigate new types of material candidates and explore the design space to optimize the performance of actuators that derives their actuation response from nano-scale and micro-scale structures. This analysis can be extended to other kinds of TCPAs such as carbon nanotubes bundles-based, hydrogen-based and the ones based on carbon fibers.
\end{itemize}

\section*{Acknowledgement}
The authors would like to thank Dr. Royan D'Mello, Dr. Shardul Panwar, and Mr. Ahmad Shaikh for their insightful discussions. 
\appendix

\section*{Appendix A: Transversely isotropic materials}\label{TransverslyIsotipropicSection}

\renewcommand{\theequation}{A.\arabic{equation}}

A general three-dimension fully anisotropic elasticity tensor has 21 independent. If there are three planes of symmetry, the number of independent constants in the elasticity tensor reduces to 9. Transverse isotropy is a special case of an orthotropic elasticity tensor with one plane of isotropy where the properties in that plane are rotation independent. A transversely isotropic elasticity tensor has 5 independent constants. One of the common examples of a transversely isotropic material is fibrous composite lamina or a laminate made up of laminae stacked in the same direction.  Hooke's law for a transversely isotropic material relates the stress and strain tensors as follows:
\begin{equation}
    \sigma_{ij} = C_{ijkl} \left(\epsilon_{kl} - \epsilon_{kl}^{th} \right) = C_{ijkl}\epsilon_{kl}^e
\end{equation}

where $C_{ijkl}$ is the fourth-order stiffness tensor of material properties or elastic moduli, $\epsilon_{kl}$, $\epsilon_{kl}^{th}$ and $\epsilon_{kl}^{e}$ are total strain tensor, thermal strain tensor and elastic strain tensor, respectively. For a transversely isotropic material, the constitutive relations takes the following form in Voigt notation with stiffness tensor bring a 6x6 tensor (in principal material direction):

\begin{equation}
\begin{bmatrix}
\sigma_{11} \\
\sigma_{22} \\
\sigma_{33} \\
\tau_{12} \\
\tau_{13} \\
\tau_{23}
\end{bmatrix}
=
\begin{bmatrix}
C_{11} & C_{12} & C_{13} & 0 & 0 & 0 \\
C_{12} & C_{11} & C_{13} & 0 & 0 & 0 \\
C_{13} & C_{13} & C_{33} & 0 & 0 & 0 \\
0 & 0 & 0 & C_{44} & 0 & 0 \\
0 & 0 & 0 & 0 & C_{44} & 0 \\
0 & 0 & 0 & 0 & 0 & (C_{11}-C_{12})/2
\end{bmatrix}
\begin{bmatrix}
\epsilon_{11} -\epsilon_{11}^{th}\\
\epsilon_{22} -\epsilon_{22}^{th}\\
\epsilon_{33} -\epsilon_{33}^{th}\\
\gamma_{12}/2 \\
\gamma_{13}/2 \\
\gamma_{23}/2
\end{bmatrix}
\end{equation}

where, $\epsilon_{ii}^{th}=\alpha_{ii}(\Delta T)$is the thermal strain component at temperature $\Delta T$ from initial state. Here $\alpha_{ii}$ is the thermal expansion coefficient in principal material direction $i$. An inverse relation can be written as:
\begin{equation}
    \epsilon_{ij}^e = S_{ijkl} \sigma_{kl}
\end{equation}

where $S_{ijkl}$ is a fourth-order compliance tensor which can further be written in terms of engineering constants of the material as follows:
\begin{equation}    
\begin{bmatrix} 
\epsilon_{11}^e\\
\epsilon_{22}^e \\ 
\epsilon_{33}^e \\
\gamma_{12}^e/2 \\
\gamma_{13}^e/2 \\
\gamma_{23}^e/2 
\end{bmatrix}=
\begin{bmatrix} 
1/E_1 & -\nu_{12}/E_1 & -\nu_{12}/E_1 & 0 & 0 & 0 \\ 
-\nu_{12}/E_1 & 1/E_1  & -\nu_{23}/E_3 & 0 & 0 & 0 \\
-\nu_{12}/E_1 & -\nu_{23}/E_3 & 1/E_3 & 0 & 0 & 0 \\
0 & 0 & 0 & 1/G_{12} & 0 & 0 \\
0 & 0 & 0 & 0 & 1/G_{13} & 0 \\
0 & 0 & 0 & 0 & 0 & 1/G_{23} 
\end{bmatrix} 
\begin{bmatrix}
\sigma_{11} \\
\sigma_{22} \\
\sigma_{33} \\
\tau_{12} \\
\tau_{13} \\
\tau_{23}
\end{bmatrix}
\end{equation}

 \section*{References}

\bibliography{reference_list}
\end{document}